\documentclass[aps,prl,preprint,superscriptaddress]{revtex4}
\usepackage{graphicx}
\usepackage[english]{babel}
\usepackage{epstopdf}
\usepackage{amsmath}

\newcommand{\C}{$^{\circ}$C}

\begin{document}
 \preprint{In press, Appl. Phys. Lett.}

 \title{Determining exciton bandwidth and film microstructure in polythiophene films using linear absorption spectroscopy.}
 
 \author{Jenny Clark}
 \affiliation{Politecnico di Milano, Dipartimento di Fisica, Piazza Leonardo da Vinci, 32, Milano 20133, Italy}
 \author{Jui-Fen Chang}
 \affiliation{Cavendish Laboratory, University of Cambridge, J.J. Thomson Avenue, Cambridge, CB3 OHE, United Kingdom}
 \author{Frank C. Spano}
 \affiliation{Department of Chemistry, Temple University, Beury Hall 201, Philadelphia, Pennsylvania 19122, United States of America}
 \author{Richard H. Friend}
  \affiliation{Cavendish Laboratory, University of Cambridge, J.J. Thomson Avenue, Cambridge, CB3 OHE, United Kingdom}
 \author{Carlos Silva}
 \affiliation{D\'{e}partement de physique et Regroupement qu\'{e}b\'{e}cois sur les mat\'{e}riaux de pointe, Universit\'{e} de Montr\'{e}al, C.P. 6128, Succursale centre-ville, Montr\'{e}al QC H3C 3J7, Canada}

 \date{\today}
 
 \begin{abstract}
We analyze the linear absorption spectrum of regioregular poly(3-hexylthiophene) films spun from a variety of solvents to probe directly the film microstructure and how it depends on processing conditions. We estimate the exciton bandwidth and the percentage of the film composed of aggregates quantitatively using a weakly interacting H-aggregate model. This provides a description of the degree and quality of crystallites within the film and is in turn correlated with thin-film field-effect transistor characteristics.
 \end{abstract}
 

\maketitle


In thin films, regioregular poly(3-hexylthiophene) (rrP3HT) self-organizes into two-dimensional $\pi$-stacked lamellar structures \cite{Mccullough_1993, Sirringhaus_1999}. The degree of order within and around these supramolecular structures depends sensitively on polymer characteristics and thin film processing conditions \citep{Kline_2006_Rev, Sirringhaus_1999}. These in turn greatly affect device performance.
Here we use linear absorption spectroscopy as a simple probe of how thin film microstructure varies with processing conditions. 
The absorption spectrum of rrP3HT thin films can be comprehensively explained using a weakly interacting H-aggregate model \cite{Clark_2007, Spano_2005_P3HT, Spano_2006_ChemPhys}. This model can be used to dissect the absorption spectrum, providing quantitative estimates not only of the fraction of the film made up of aggregates (recently determined using different methods \cite{Moule_2008, Berson_2007}), but also the degree of excitonic coupling within the aggregates, a parameter which is related to average conjugation length and the crystalline quality. Using this information, we obtain an overall description of the microstructure of the films, which we find to be well correlated with thin-film field effect transistor (FET) characteristics.

rrP3HT (Plextronics and Merck) was dissolved in various solvents (Aldrich) at a concentration of 10\,gL$^{-1}$ and spin-coated onto either fused silica spectrocil B or patterned FET substrates. The films were annealed at 100\C~for 1\,hr. The FET-patterned substrates were fabricated in a bottom gate, bottom-contact configuration. The gate electrode was a heavily n-doped Si wafer with 200\,nm thermally grown SiO$_2$ as dielectric. A Au layer of 15--20\,nm was patterned on the substrates as source/drain electrodes with a channel length of 20\,$\mu$m by conventional photolithography. The channel widths were 1 cm. The patterned substrates were treated with hexamethyldisilazane (HMDS). All processing, and I-V measurements were performed in a N$_{2}$ atmosphere. Absorption measurements were performed in ambient conditions using a HP 8453 uv-vis spectrometer and tapping-mode atomic force microscopy (AFM) measurements were taken using a Digital Instruments Dimension 3100 AFM.

\begin{figure}[ht]
	\centering
	\includegraphics{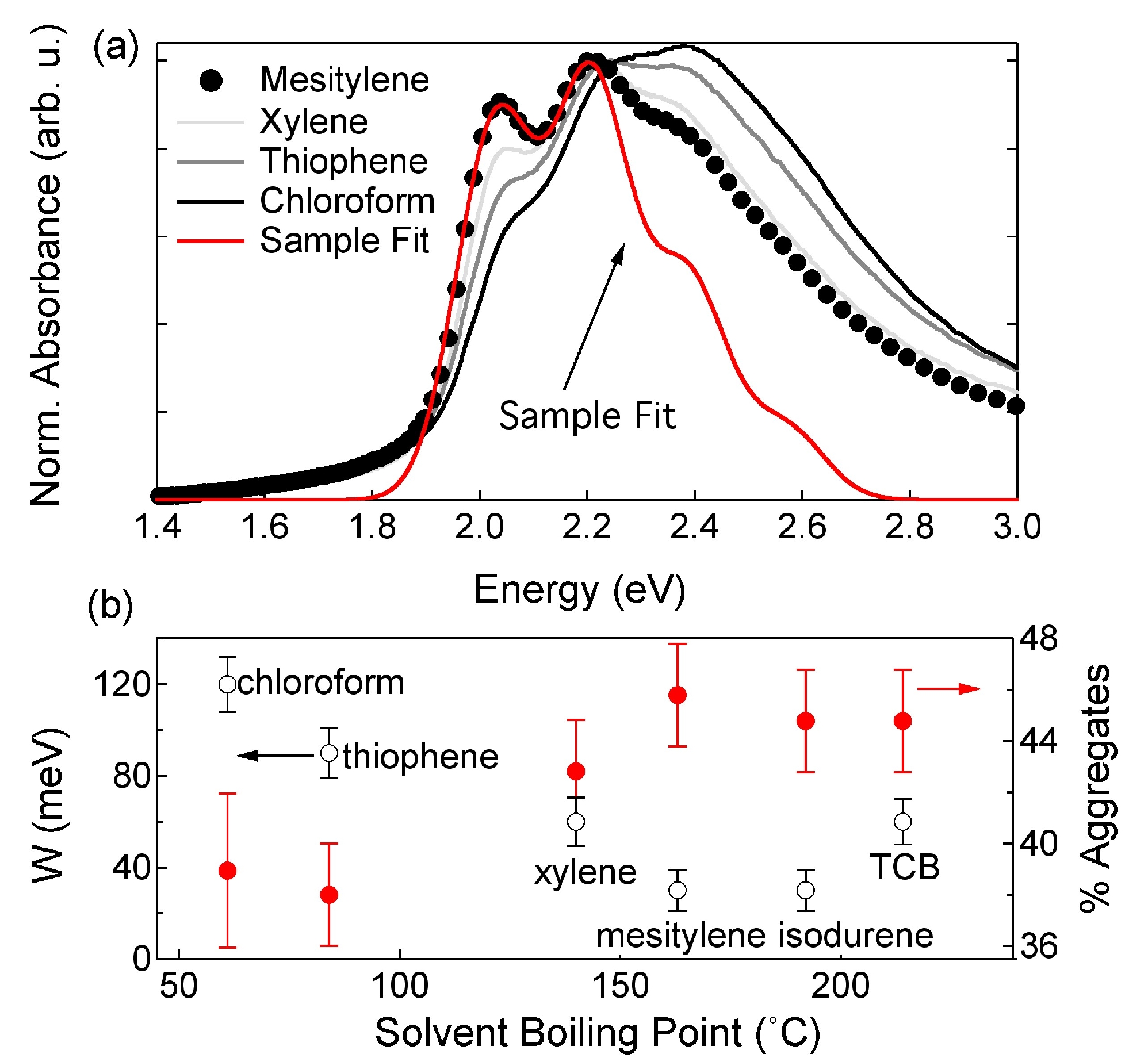}
	\caption{(a) Normalized Absorbance spectra taken on films spun from different solvents. A sample fit is also shown. (b) Exciton bandwidth, $W$ (open black circles), left axis and percentage of film made up of aggregates (full red circles), right axis. (TCB is 1,2,4 tri-chlorobenzene). }
	\label{Fig1}
\end{figure}

The absorbance spectra of rrP3HT thin films spun from various solvents are shown in Fig.~\ref{Fig1}\,(a). 
The film absorption spectrum is composed of two parts \cite{Clark_2007}, a lower energy, dominant part from crystalline regions of the film which form weakly interacting H-aggregate states and a higher energy part due to more disordered chains which form \emph{intrachain} states. The ratio of the 0--0 and 0--1 peak absorbance is related to the exciton bandwidth of the aggregates, $W$, and the energy of the main oscillator coupled to the electronic transition, $E_{p}$, by the following expression (assuming a Huang-Rhys factor of 1 \cite{Spano_2005_P3HT, Clark_2007}):
 \begin{equation}
		\frac{A_{0-0}}{A_{0-1}} \approx \left(\frac{1-0.24W/E_{p}}{1+0.073W/E_{p}} \right)^2.
		\label{Spano_Equation2b}
\end{equation}
Using the 0--0/0--1 ratio from Fig.\ \ref{Fig1}, and assuming the C=C symmetric stretch at 0.18\,eV dominates the coupling to the electronic transition \cite{Louarn_1996}, $W$ can be estimated. Assuming similar interchain order in the films, $W$ is related to the conjugation length and \emph{intra}chain order. An \emph{increase} in conjugation length and order will lead to a \emph{decrease} in $W$ \citep{Manas_1998, Beljonne_2000, Barford_2007}.
$W$ is shown in part (b) of Fig.\ \ref{Fig1} as a function of solvent boiling point (bp). Also shown in the figure is the percentage of the film made up of aggregates. This is found by determining first the relative oscillator strength of the aggregated species compared with the intrachain species and second, the percentage of the absorption spectrum made up of aggregate absorption. 

The relative oscillator strength of the aggregated versus intrachain species is determined by examining the absorption spectrum of a 1\%\,wt solution as a function of temperature (Fig.\ 1 of Ref.\ \cite{Clark_2007}). The isosbestic point demonstrates that the reduction in the number of chains in the amorphous phase is equal to the increase in the number of chains in the aggregated phase. Therefore, the increase in oscillator strength going from the amorphous to the aggregated phase can be calculated (see supplementary information) and is found to be $39\pm10\%$.

The fraction of the spectrum made up of aggregate absorption is determined using a modified Frank-Condon fit to the absorption spectrum, which takes into account the H-aggregate nature of the state (Ref. \cite{Spano_2006_ChemPhys}),
\begin{equation}\label{Spano_Abs_Fit}
A\propto \sum_{m=0}\left(\frac{e^{-S}S^m}{m!}\right)\left(1-\frac{We^{-S}}{2E_p}G_m\right)^2 \Gamma(\hbar \omega -E_{0-0} - mE_{p}).
\end{equation}
The vibrational level is denoted $m$ and $G_m$ is a constant that depends on $m$ according to $G_m=\Sigma_{n (\neq m)}\lambda^{2n}/n!(n-m)$, where the sum is over vibrational states, $n$ \cite{Spano_2006_ChemPhys}. For simplicity we used a Gaussian lineshape, $\Gamma$, with the same width for each vibronic transition. Fig.\ \ref{Fig1}(a) shows a sample fit for the film spun from mesitylene. Note that in Eq.~\ref{Spano_Abs_Fit} we have omitted second order corrections to the vibrational peak frequencies \cite{Spano_2006_ChemPhys}.

As a general trend, Fig.~\ref{Fig1}~(b) demonstrates that films spun from low bp solvents, such as chloroform, exhibit larger $W$ and hence shorter conjugation lengths (lower crystalline quality), while those from higher bp solvents show lower $W$ with correspondingly longer conjugation lengths (higher crystalline quality). Furthermore, films spun from low bp solvents have a lower proportion of crystalline aggregated regions compared with those spun from high bp solvents, as expected \cite{Chang_2000}. 

\begin{figure}[ht]
	\centering
	\includegraphics[width=8.5cm]{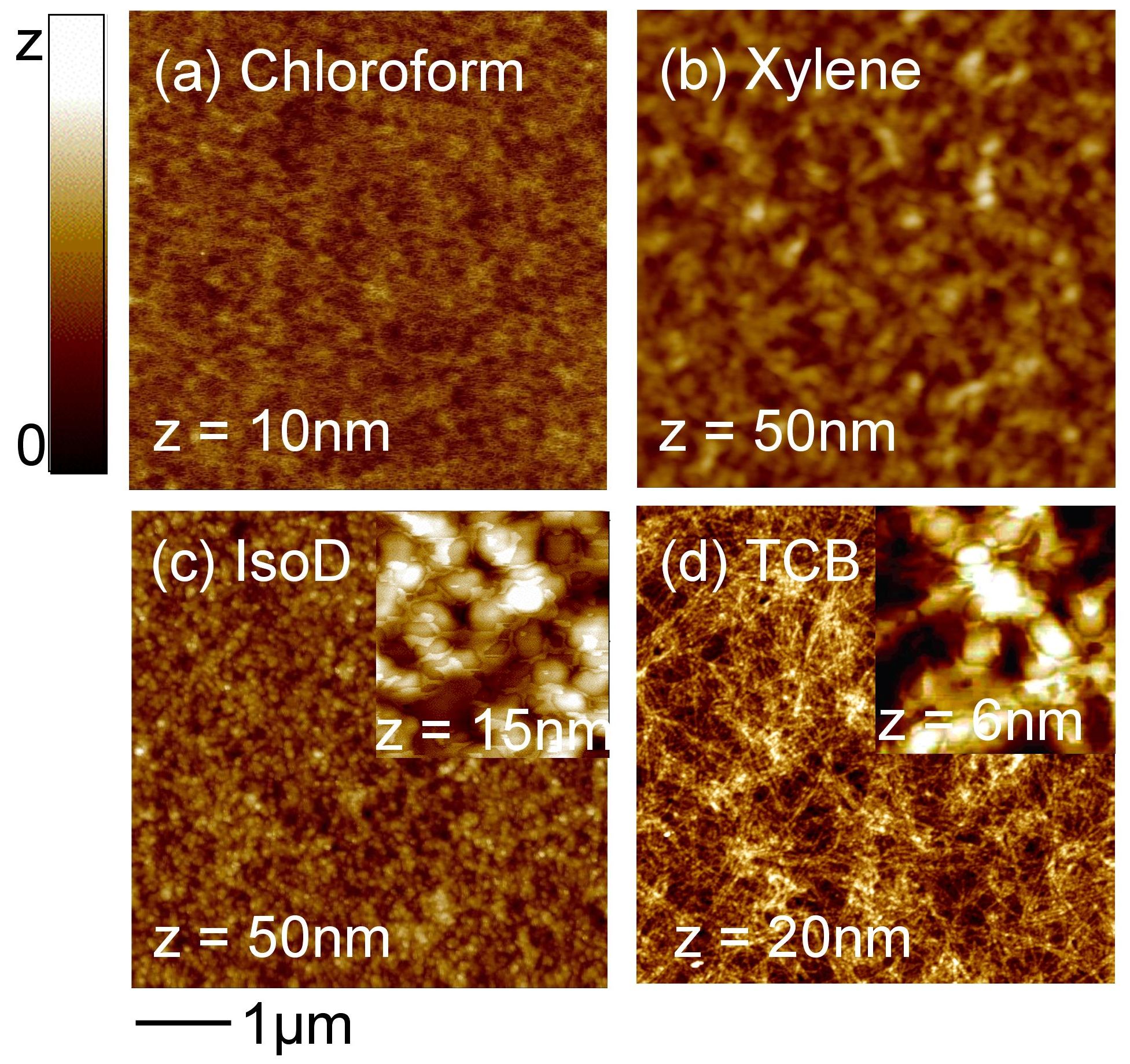}
	\caption{Tapping mode atomic force microscopy (AFM) images of films spun from (a) chloroform, (b) xylene, (c) isodurene (IsoD) and (d) 1,2,4 tri-chlorobenzene (TCB). Images are 5$\times$5\,$\mu$m. Insets are 0.5$\times$0.5\,$\mu$m. Height-scales (z) are marked in the figure.}
	\label{Fig2}
\end{figure}

To verify that the changes in the spectra are microstructural in origin, we have studied the morphology of the films using AFM.
Fig.\ \ref{Fig2} shows AFM images of films spun from various solvents. The films spun from high bp solvents show rough surfaces (isodurene film root-mean-square (rms) roughness 5.7\,nm), while those spun from low bp solvents are smoother (chloroform film rms roughness 0.95\,nm).
The roughness is due to the formation of micro-structures \cite{Chang_2000, Kline_2006_Rev, Chang_2004, Berson_2007, Moule_2008}. 

We therefore find that films with increased average conjugation lengths, order and percentage of crystalline structures, as predicted from the absorption spectra, demonstrate morphologies with micro-crystalline structures. Those with an increased percentage of disordered regions in the film demonstrate amorphous films. The absorption spectrum is therefore a useful and easy-to-measure tool to determine film micro-structure.

\begin{figure}[htb]
	\centering
	\includegraphics[width=8.5cm]{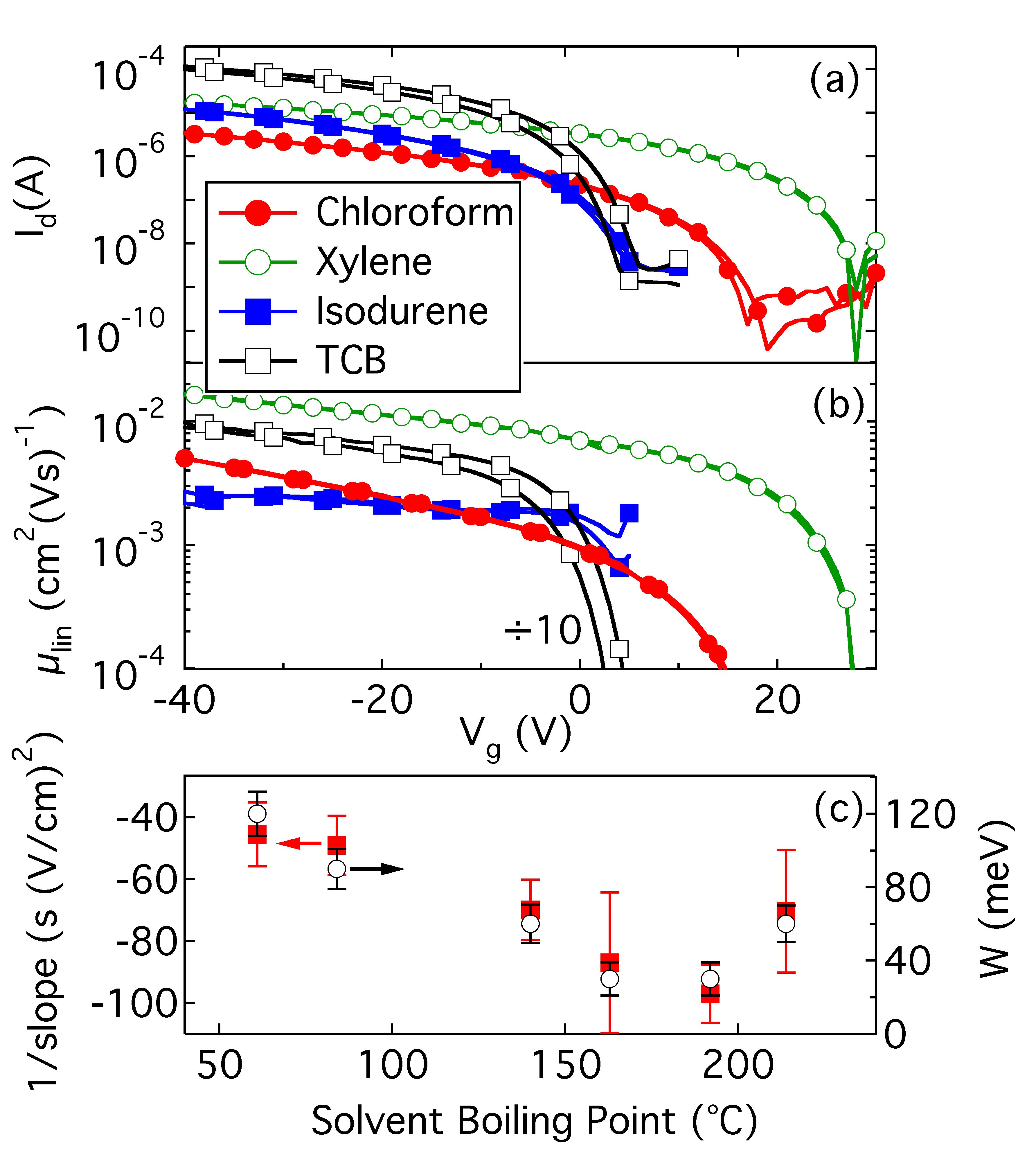}
	\caption{Linear regime ($V_{ds}$=-5\,V) (a) transfer characteristics and (b) mobility as a function of gate voltage for films spun from different solvents (TCB mobility is divided by 10 for easier comparison). (c) Inverse slope of mobility versus V$_{g}$ (left axis, red squares) and free exciton bandwidth, $W$ (right axis, black open circles) as a function of solvent boiling point.}
	\label{Fig3}
\end{figure}

Transfer curves and plots of mobility versus gate-voltage (V$_{g}$) of transistors spun from a variety of solvents are shown in Fig.s~\ref{Fig3}~(a) and (b), respectively. Using a linear fit of the forward sweep of the mobility, from V$_{g}$=-10 to -40\,V, we estimated how the slope of the mobility versus V$_{g}$ changes as a function of solvent bp. This is shown in Fig.~\ref{Fig3}~(c) (left axis). Also plotted is the free exciton bandwidth, $W$ (right axis), reproduced from Fig.~\ref{Fig1}~(c). There is an excellent correlation between $W$, obtained from the absorbance spectra, and the degree to which the mobility is V$_g$-dependent. Changes in slope of the mobility versus V$_g$ as a function of solvent bp have previously been noted \cite{Chang_2004} and have been explained in the context of the multiple trapping and release model \cite{Horowitz_1999, Sirringhaus_1998, Salleo_2004, Sirringhaus_2005}. The model assumes a large concentration of localized states below the mobility edge. At low V$_g$, most of the charges fill the localized states (traps) and the mobility is low. As V$_g$ increases more of the traps are filled as the Fermi level moves closer to the mobility edge. More charges can be released to the extended, mobile states which leads to an increase in the effective mobility. At sufficiently high V$_g$, all of the traps are filled and any additional charges are injected into the extended states. Once this occurs, the mobility is constant with V$_g$. The degree to which the mobility is a function of V$_g$ is thus dependent on the width of the localized density of states below the mobility edge. Shallow trap distributions will result in a weakly V$_g$-dependent mobility, while wider trap distributions will result in a pronounced V$_g$-dependence. 
From Fig.\ \ref{Fig3} we can therefore deduce that, for example, the device spun from isodurene has a shallow distribution of traps, while that spun from chloroform has a wide trap distribution. 
There is a clear correlation between the crystalline quality, \emph{as determined from the absorbance spectrum}, and the width of the trap distribution.

We note however, that if we assume that there is no change in spatial disorder between hopping sites, we would expect devices with wider trap distributions to have correspondingly lower mobilities. Fig.\ \ref{Fig3}~(b) shows that this is not the case. The reason for this is unclear from our absorption spectra analysis, probably due to the fact that the absorbance spectra provides a measure of the bulk film microstructure, while properties at interfaces are often important, especially in FETs where transport occurs in a narrow region close to the dielectric/semiconductor interface \cite{Chang_2006, Kline_2006_NMat}. 

Nevertheless, we have demonstrated that the absorbance spectrum by itself, can be used as a simple yet powerful probe of thin film excitonic coupling, intrachain order and fraction of crystalline regions within the film. Furthermore, it can be used to predict the relative distribution of trap states below the mobility edge in FETs, providing evidence that the quality of the crystallites is homogeneous throughout the film and does not change between the bulk and the interface. 
Therefore, linear absorption spectroscopy is a usefully simple tool to determine bulk film microstructure and this knowledge will help understanding of device characteristics in a variety of optoelectronic devices \cite{Moule_2008}.

\begin{acknowledgments}
JC acknowledges support from EPSRC and an industrial Case award from Seiko Epson UK. CS acknowledges support from the NSERC and the Canada Research Chairs program. FCS is supported by the NSF, Grant No. DMR 0606028. 
\end{acknowledgments}

\newpage

\end{document}